\documentclass{iaus}
\usepackage{graphicx}

\title[Synthetic Milky Way emission maps] 
{Using synthetic emission maps to constrain the structure of the Milky Way}

\author[A. R. Pettitt \it{et al. }]   
{Alex R. Pettitt$^1$, Clare L. Dobbs$^1$, \\ David M. Acreman$^1$ \and Daniel J. Price$^2$}

\affiliation{$^1$ School of Physics \& Astronomy, University of Exeter, Stocker Road, Exeter EX4 4QL\\ email: {\tt alex@astro.ex.ac.uk} \\[\affilskip]
$^2$ Monash Centre for Astrophysics (MoCA), School of Mathematical Sciences, Monash University, Vic. 3800, Australia }

\pubyear{2014}
\volume{298}  
\pagerange{XX -- YY}
\jname{IAUS\,298 Setting the scence for Gaia and LAMOST}
\editors{S. Feltzing, G. Zhao, N.\,A. Walton \& P.\,A. Whitelock, eds.}
\begin{document}

\maketitle

\begin{abstract}
We present the current standing of an investigation into the structure of the Milky Way. We use smoothed particle hydrodynamics (SPH) to simulate the ISM gas in the Milky Way under the effect of a number of different gravitational potentials representing the spiral arms and nuclear bars, both fixed and time-dependent. The gas is subject to ISM cooling and chemistry, enabling us to track the CO and HI density. We use a 3D grid-based radiative transfer code to simulate the emission from the SPH output, allowing for the construction of synthetic longitude-velocity maps as viewed from the Earth. By comparing these maps with the observed emission in CO and HI from the Milky Way (\cite[Dame et al. 2001, Kalberla et al. 2005]{Dame_etal01, Kalberla_etal05}), we can infer the arm/bar geometry that provides a best fit to our Galaxy. By doing so we aim to answer key questions concerning the morphology of the Milky Way such as the number of the spiral arms, the pattern speeds of the bar(s) and arms, the pitch angle of the arms and shape of the bar(s).
\end{abstract}

\firstsection 
\section{Introduction}
Whilst we are able to observe arm and bar structures in nearby galaxies, we remain ignorant as to the structure of our own Galaxy, due to our unique position in the Galactic disk. Calculating the distances to sources in the Galactic disk from their velocities is the primary way of mapping the Galaxy's face-on structure, but it is fraught with difficulties and uncertainties, such as the distance ambiguity. As such we cannot say with much certainty even how many spiral arms our Galaxy has (see \cite[Vall{\'e}e 2008]{Vallee08} and references therein).

\begin{figure}
\begin{center}
 \includegraphics[trim=0cm 1cm 0cm 1.0cm,width=2.6in]{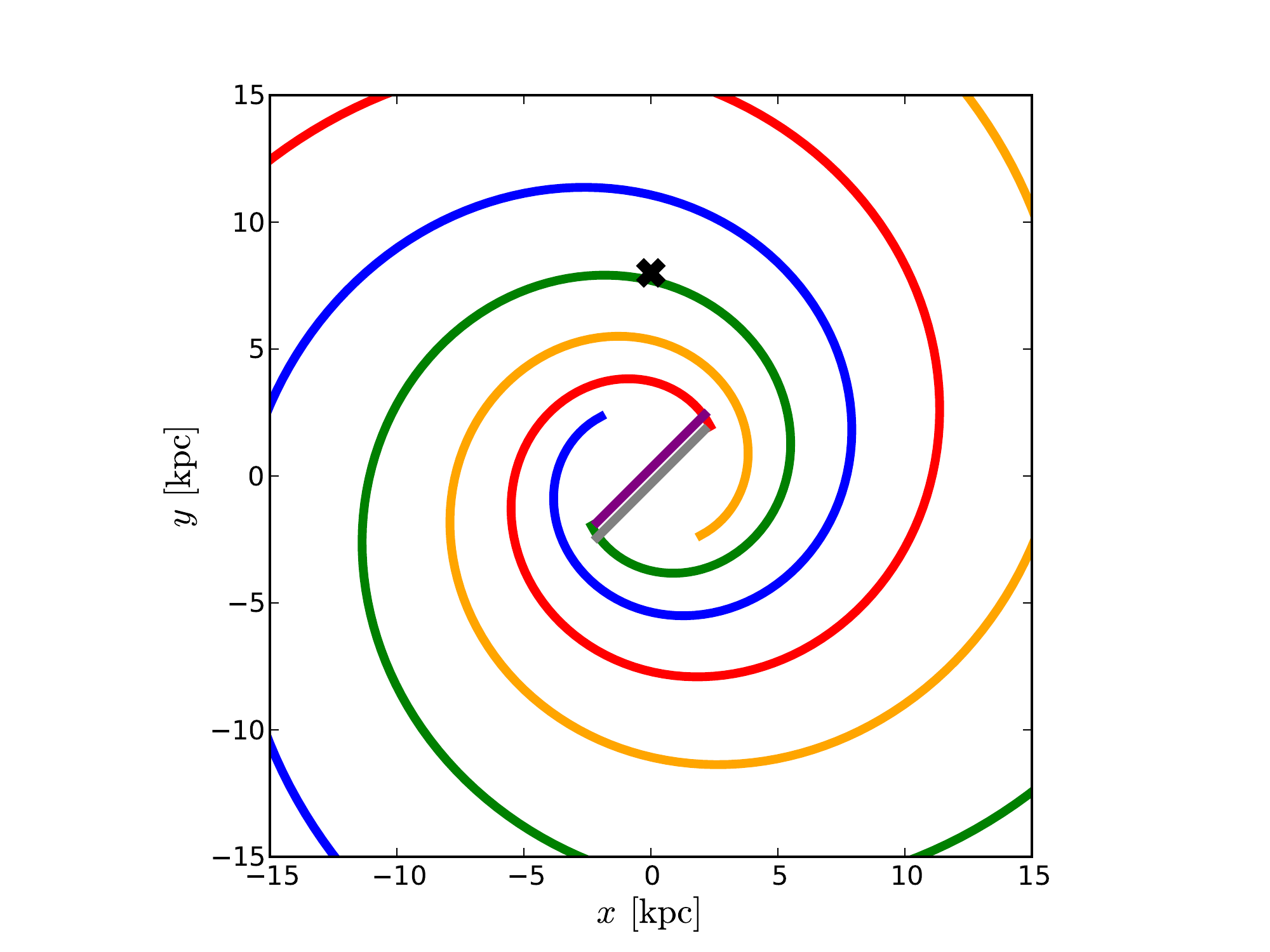} 
 \includegraphics[trim=0cm 1cm 0cm 1.0cm,width=2.6in]{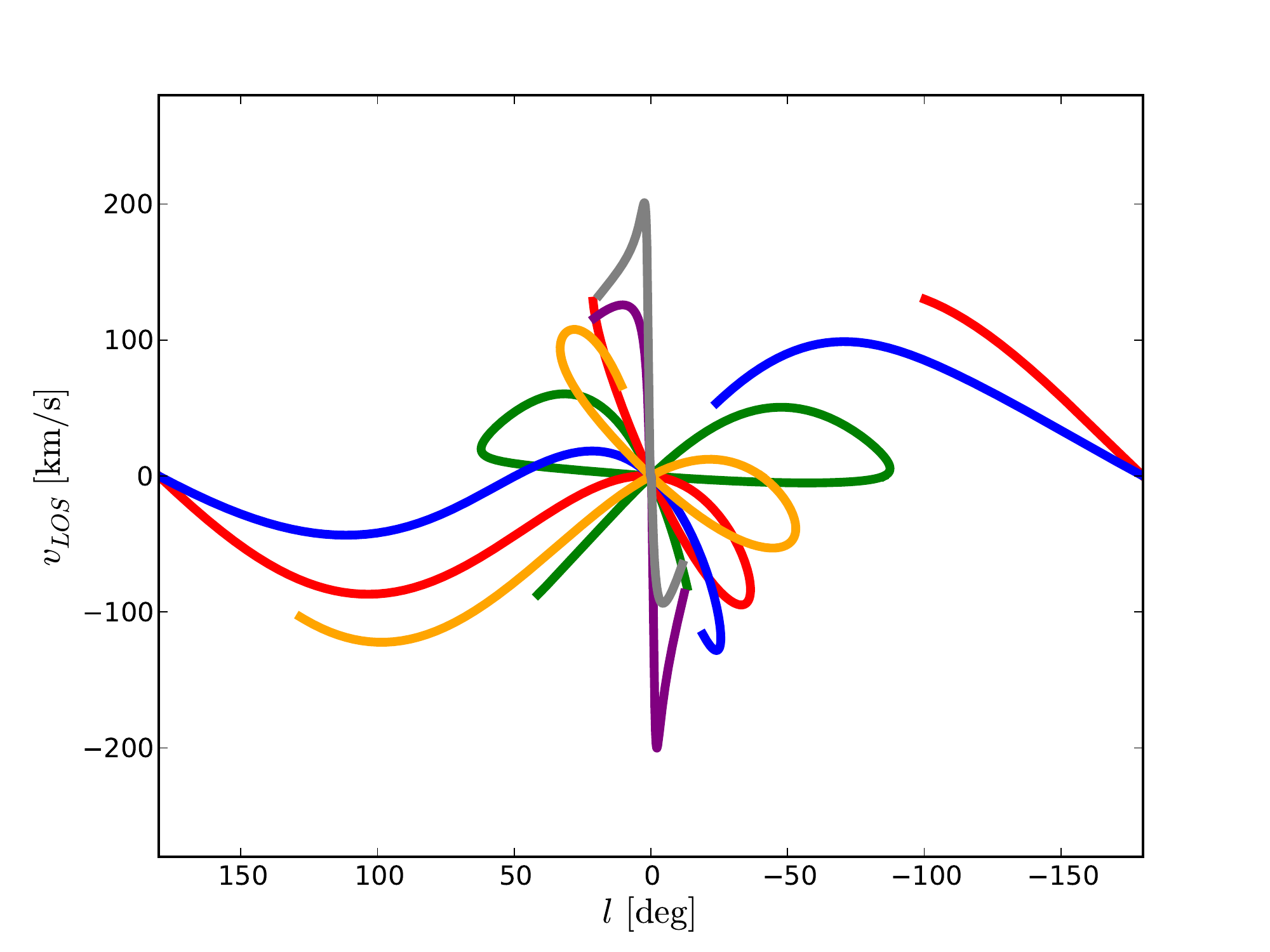} 
 \caption{Demonstration of the translation of a barred-spiral pattern in Cartesian coordinates (left) to equivalent plot in longitude-velocity space (right) as viewed by an observer (black cross, left), assuming a Milky Way-like rotation curve and a distance of 7.5\,kpc from the Galactic centre.}
   \label{ARP:fig1}
\end{center}
\end{figure}

One way of avoiding the problems associated with converting velocities to distances is to map the velocity distribution of Galactic sources as a function of Galactic longitude. An illustration of the translation from spiral and bar positions in Cartesian space into longitude-velocity (\emph{l-v}) space is shown in Fig. \ref{ARP:fig1}. By knowing the position of the observer and assuming some rotation curve we can create maps in \emph{l-v} space. In reality these features will be combined with emission from the Galactic disk and the features may not directly translate from position to velocity space. For example, spiral shocks may cause the high density gas to be offset from the actual spiral perturbation.

There exists in the literature numerous studies of the velocity structure of the Galactic disk from different sources (e.g. HI, CII, CO, masers, HCN). Of key interest is the velocity structure of HI and CO gas. While HI is believed to be present throughout the galactic disk, CO is a tracer of high density regions that coincide with the location of spiral and bar structure (\cite[Kalberla et al. 2005, Dame et al. 2001]{Kalberla_etal05,Dame_etal0}). Using these two gases we could construct a top-down map of our Galaxy, including the positions and shapes of spiral and bar features, but as mentioned converting from velocity to distance is difficult.

One approach to inferring the structure of the Galaxy from \emph{l-v} maps is to use hydrodynamical simulations to investigate which spiral and bar structures can reproduce the velocity structures seen in observations. Velocity maps built from simulations provide all the spatial information of the gas, as well as its velocity. If it is possible to reproduce the observed \emph{l-v} features in simulations then it can be inferred that the spiral/bar structures that produced these features are a good representation of those of our Galaxy. This technique has been used in previous studies in an attempt to constrain specific Milky Way parameters, such as the bar's orientation (e.g. \cite[Rodriguez-Fernandez \& Combes, G{\'o}mez \& Cox 2004]{RFRC08,GomezCox08}). So far, however, no work has been attempted to search a large parameter space of Galactic spiral and bar features. The aim of the work presented here is to match the CO and HI \emph{l-v} maps of our Galaxy by running numerous hydrodynamical simulations of gases in the interstellar medium (ISM) under the influence of spiral and bar perturbations with various pattern speeds, pitch angles, and orientations. A major difference to previous studies is that we construct synthetic \emph{emission maps} of HI and CO rather than simply translating the positions in the hydrodynamical simulations into \emph{l-v} space (as done in Fig. \ref{ARP:fig1}).

\section{Galactic simulations}

We use smoothed particle hydrodynamics (SPH) to simulate the flow of ISM gas in the Milky Way. The ISM gas is distributed in the Galactic disk only, with a distribution that matches the surface density profile of gas observed in the Milky Way. Each SPH particle has a chemical abundance array that is updated along with the various hydrodynamical properties. Our HI and H$_2$ chemistry is described in \cite[Dobbs et al. (2008)]{Dobbs_etal08}. In order to construct molecular \emph{l-v} emission maps we also include CO chemistry. We use the CO rate equations of \cite[Nelson \& Langer (1997)]{NelsonLanger97} that treats the CII to CO conversion as a single step process. The chemistry changes on a timescale much shorter than the dynamical time. As such we sub-cycle the chemistry inside the hydro time-steps. Our gas is also subject to ISM heating and cooling, from  \cite[Glover \& Mac Low (2007)]{GloverMLow_07}, see \cite[Dobbs et al. (2008)]{Dobbs_etal08}.

For the majority of our simulations we use fixed analytic potentials to represent the stellar mass distribution using the SPH code \textsc{phantom} \cite[(Price \& Federrath 2010, Lodato \& Price 2010)]{PriceFed10,LodatoPrice10}. \textsc{phantom} is a low-memory, highly efficient SPH code written especially for studying non-self-gravitating problems. When using fixed analytic potentials the structure of the Milky Way is assumed to be that of a grand design, with the gas clearly tracing the shape of the potentials. The rotation curve of the Milky Way is reproduced using a combination of bulge, disk and halo potentials. We include several different potentials to represent the spiral arms (\cite[Cox \& G{\'o}mez 2002]{Cox02}, \cite[Martos et al. 2005]{Martos_etal05}) and bars (\cite[{Long} \& {Murali} 1992, Dehnen 2000, Wada \& Koda 2001, Wang et al. 2012]{LongMurali92, Dehen00, WadaKoda01, Wang_etal12}). While we have many different potentials for each structure, each serves a separate purpose. For example, the bar of \cite[Wang et al. (2012)]{Wang_etal12} has been tailored to match the boxy/peanut density profile seen in observations while the arms of \cite[Martos et al. (2005)]{Martos_etal05} are capable of producing 4-armed spirals from only imposing a 2-armed potential.

\begin{figure}
\begin{center}
\includegraphics[trim=0cm 10.5cm 0cm 1.5cm,width=5.2in]{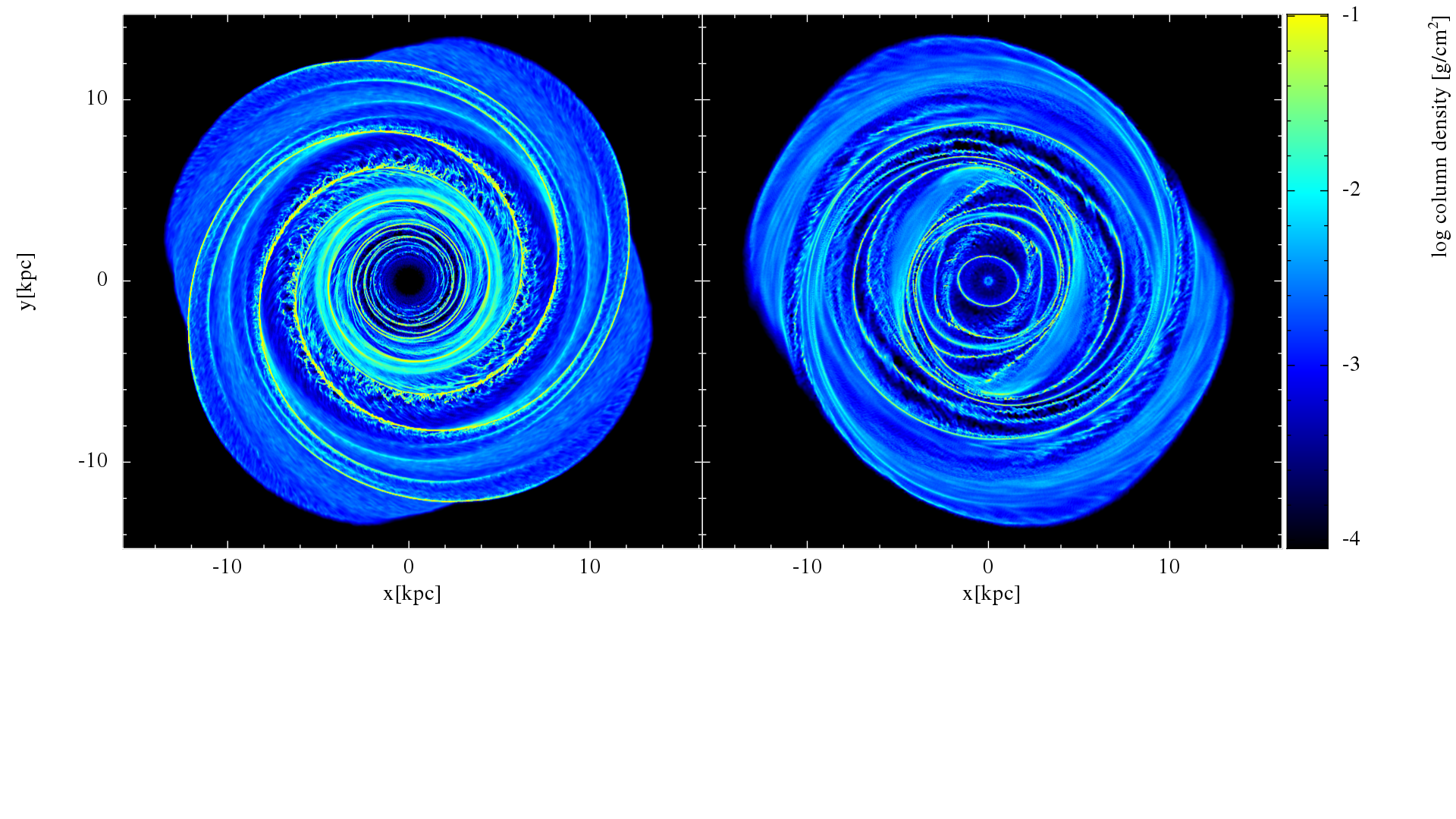} 
\caption{Galactic simulations using static analytic potentials for the stellar component. Left: a 4-armed spiral potential with a pattern speed of $20\,\rm km\, s^{-1} \, kpc^{-1}$ and pitch angle of 18$^\circ$. Right: a 2-armed spiral potential with the same pattern speed and pitch angle as the simulation on the left but with the inclusion of a central bar potential. The bar potential is from \cite[{Long} \& {Murali} (1992)]{LongMurali92} with dimensions of 4:1:1\,kpc in x:y:z and a pattern speed $40\,\rm km\, s^{-1} \, kpc^{-1}$.}
\label{ARP:fig2}
\end{center}
\end{figure}

Two example simulations are shown in Fig. \ref{ARP:fig2}. On the left is a grand design 4-armed spiral galaxy. The primary arms are clearly visible, along with several weaker interarm features. The simulation on the right is of a 2-armed barred spiral, where the bar has dimensions of 4:1:1\,kpc in x:y:z. The spiral perturbations in both simulations are moving with a pattern speed of $20\,\rm km\, s^{-1} \, kpc^{-1}$ and the bar is rotating at $40\,\rm km\, s^{-1} \, kpc^{-1}$. The bar is angled $20^\circ$ from the y-axis. The addition of a bar creates a plethora of extra structure, and this is seen regardless of the exact form of the bar potential used.

We can also represent the stellar matter in the Milky Way as SPH particles themselves. The gravitational attraction of the star particles is felt by the star and gas particles, but only the gas is subject to the standard hydrodynamical forces. The calculations using live star particles were performed using the SPH code \textsc{sphNG} based on the original version of \cite[Benz et al. (1990)]{Benz_etal90}, but substantially modified as described in \cite[Bate et al. (1995)]{Bate_etal95} and \cite[Price \& Monaghan (2007)]{PriceMon07}. We set up the initial stellar velocities using the method of \cite[Hernquist (1993)]{Hernquist93}. The separate disk-bulge-halo components are represented by an exponential disk, a Plummer bulge and NFW halo. Our stellar particles are split between a disk and bulge population, where the former are given circular orbits of the order $200\,\rm km\,s^{-1}$ and the orbits of the latter are given random orientations. The halo component is represented by an analytic potential. Our set-up is based on that of \cite[Baba et al. (2010)]{Baba_etal2010} with the addition of a bulge component to better match the observed rotation curve of the Milky Way and reproduce the velocities in the \emph{l-v} diagram near $l=0^\circ$. For the simulations shown here 90\% of the SPH particles are allocated to the gaseous disk, 9\% to the stellar disk and 1\% to the stellar bulge.

\begin{figure}
\begin{center}
\includegraphics[trim=0cm 10.5cm 0cm 1.5cm,width=5.2in]{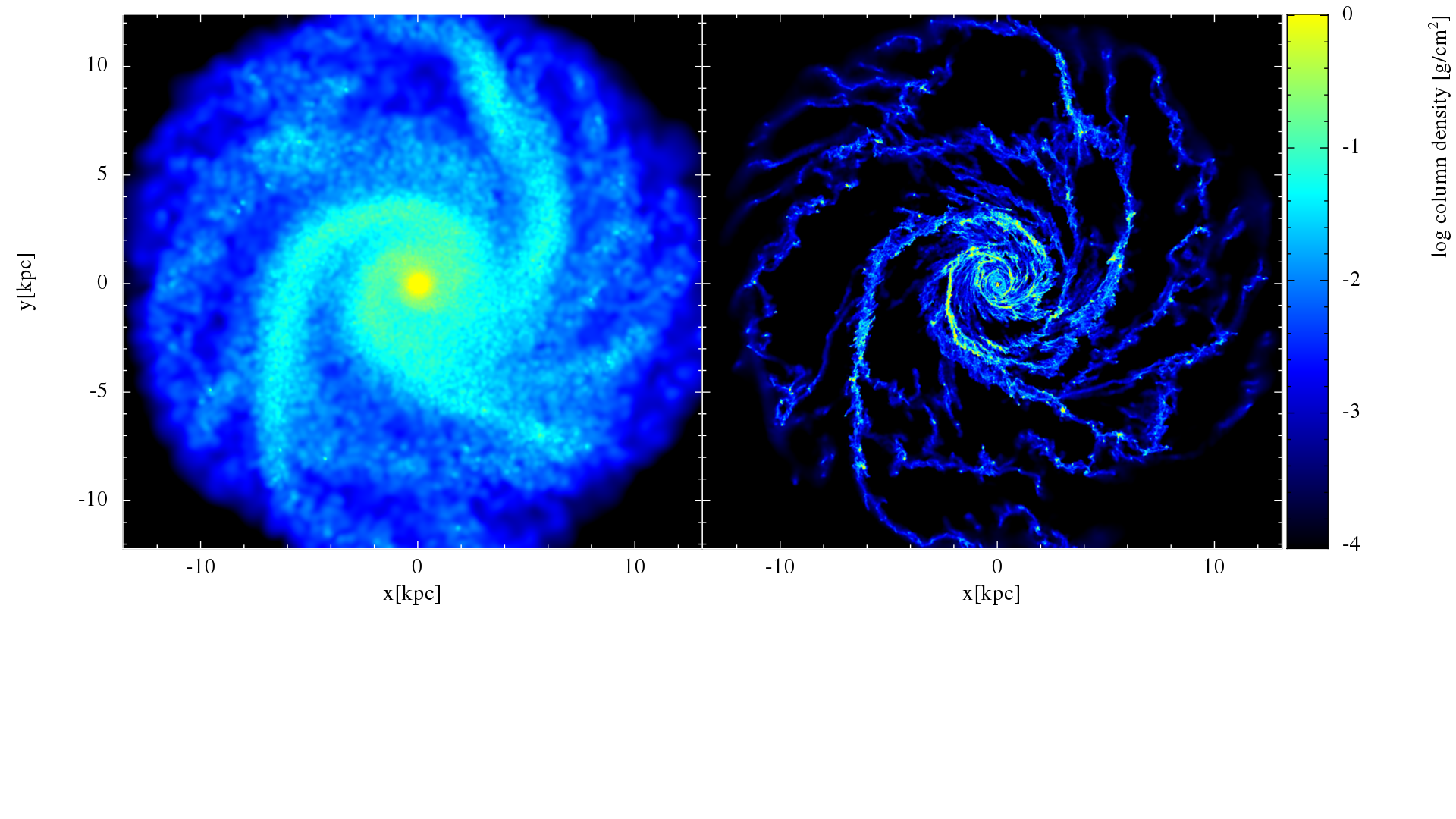} 
\caption{Simulations with a live stellar bulge and disk embedded in a static halo. Left: star particles that populate the disk and bulge, right: gas particles that  populate just the disk. The gas traces the regions of high stellar density, and in this simulation displays a strong 2/3 armed structure with numerous interarm features.}
\label{ARP:fig3}
\end{center}
\end{figure}

In Fig. \ref{ARP:fig3} we show an example of a simulation with a live stellar disk and bulge. This setup is more akin to a flocculent spiral galaxy. The live nature of the stellar component results in spiral structures that appear much weaker, transient and irregular compared to those imposed by analytic stellar potentials. While the number of primary arms tends to be steady throughout the simulation, the star particles inhabit spiral arms throughout the existence of the arms, as opposed to the density wave theory where material continually flows in and out of the spiral density waves.

\section{Emission maps}
\begin{figure}
\begin{center}
\includegraphics[trim=0cm 1cm 0cm 0cm,width=5.2in]{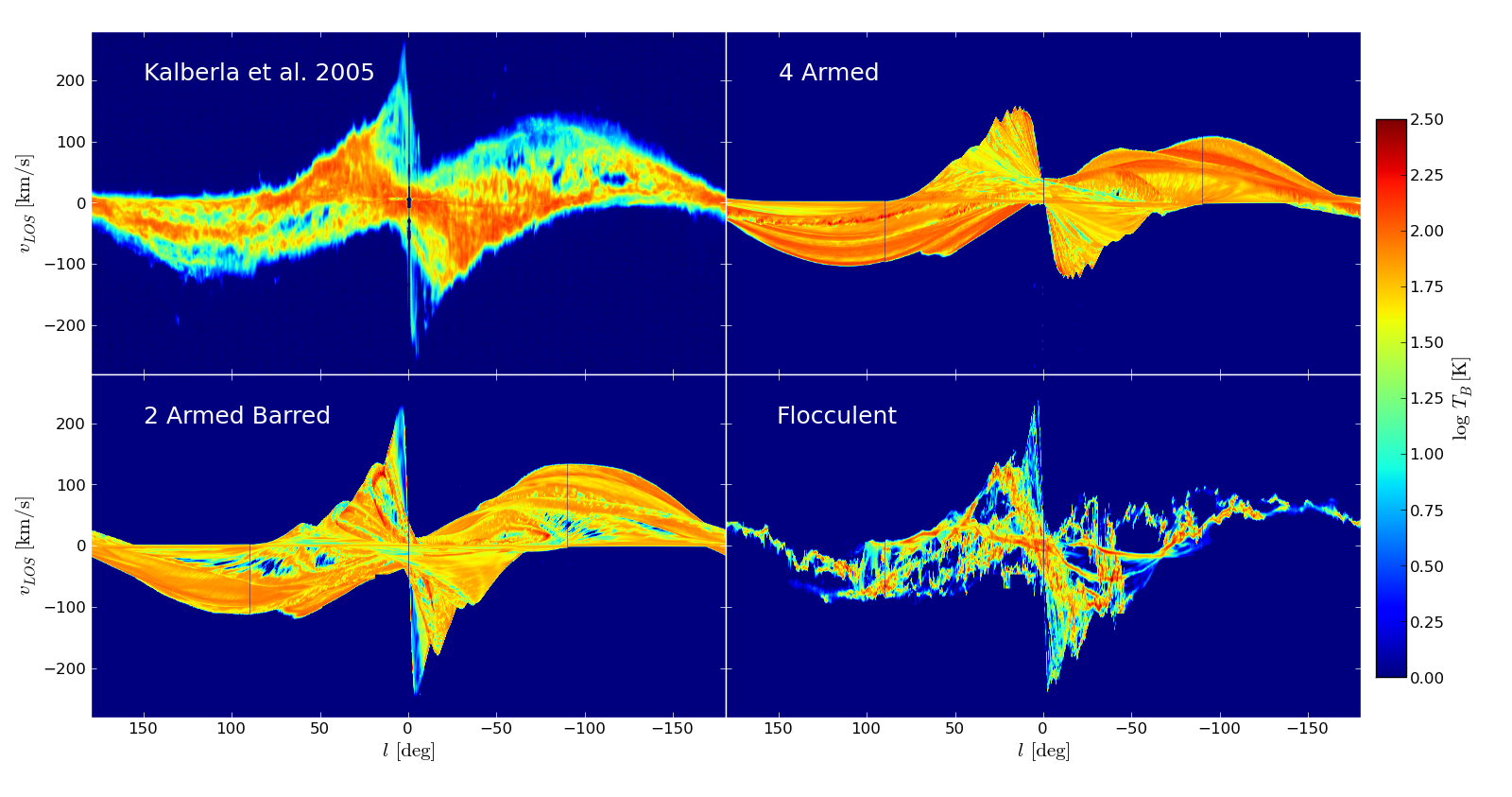} 
\caption{Synthetic emission maps of HI (21-cm line) seen from a position of 7.5\,kpc in the simulations shown in Figs \ref{ARP:fig2} and \ref{ARP:fig3}. The observed emission is shown in the top left \cite[(Kalberla et al. 2005)]{Kalberla_etal05}. The emission is shown at a cut in latitude in the Galactic plane ($b=0^{\circ}$).}
\label{ARP:fig4}
\end{center}
\end{figure}

\begin{figure}
\begin{center}
\includegraphics[trim=0cm 1cm 0cm 1.0cm, width=5.2in]{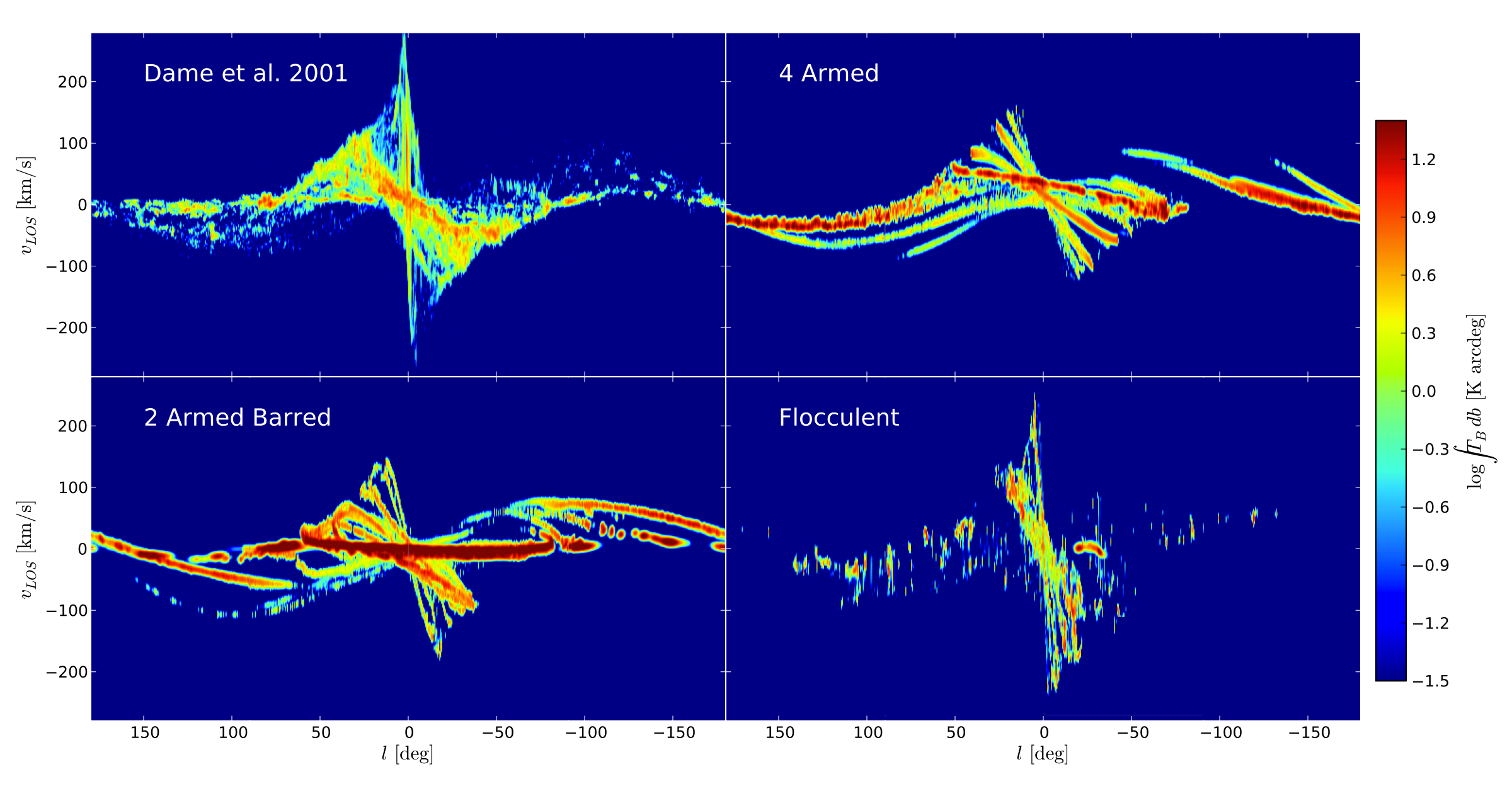}
\caption{Synthetic emission maps of CO (J=0-1 transition) seen from a position of 7.5\,kpc in the simulations shown in Figs \ref{ARP:fig2} and \ref{ARP:fig3}. The observed emission is shown in the top left \cite[(Dame et al. 2001)]{Dame_etal01}. The emission has been integrated through $-2^\circ \le b \le +2 ^\circ$. A turbulent velocity dispersion term of $\rm 4\,km\,s^{-1}$ has been added to the line width as the emission was far too sparse compared to observations.}
\label{ARP:fig5}
\end{center}
\end{figure}

Once the simulations have reached a dynamical steady state we then follow a prescription similar to that of \cite[Acreman et al. (2012)]{Acreman_etal12} to construct synthetic emission maps. We utilise a 3D grid based radiative transfer code, \textsc{torus} \cite[(Harries 2000)]{Harries00}, to calculate the emission from the HI 21-cm and CO (J=0-1) transitions. The SPH data are first interpolated to an AMR grid as described in \cite[Acreman et al. (2010)]{Acreman_etal10}. The observer is placed at a position of 7.5\,kpc from the Galactic centre. The azimuthal position of the observer is used to orientate the arm/bar features to positions suggested by observations. The emission maps for the simulations shown in Figs \ref{ARP:fig2} (grand design) and \ref{ARP:fig3} (flocculent) are shown in Figs \ref{ARP:fig4} (HI) and \ref{ARP:fig5} (CO).

The synthetic emission maps of HI reproduce the global structure of the observed emission well. The barred galaxy in particular is a good match to the broad structure and peak velocities towards the galactic centre. The flocculent galaxy is lacking in the broad structure inside $|l|<50^\circ$, however the resolution for the live disk/bulge simulations is somewhat lower than those with static disk/bulge potentials. Conversely the HI of the 4-armed and barred galaxy is tracing the spiral/bar structure too clearly compared to observations.

Our synthetic CO emission maps suffer the same problems as the HI maps, in that the emission is strongly confined to the spiral and bar structures. There is very little interarm emission in the grand design simulations and neither the 4-armed or the barred galaxy can reproduce the high velocities observed at $l\approx 0^\circ$. The arms of the flocculent galaxy are weak in CO emission compared those of the grand design, however the high velocities towards the galactic centre are present. The broad parallelogram-like structure is present but is angled too steeply in \emph{l-v} space.

\section{Conclusions}
We present the first synthetic longitude-velocity emission maps of the Milky Way's ISM tracers of structure. Our numerical simulations display a multitude of features, with arm and bar potentials driving supplementary structures visible in the emission maps. In cases where we use fixed potentials the emission features appear too strong and numerous. Arms produced in flocculent galactic simulations appear too irregular compared to the Galaxy's observed emission, but are a closer match to the observed contrast between arm and interarm emission in HI. The remainder of this work will involve a detailed search through  arm-bar parameter space to determine the morphology that best matches the structure in the observed \emph{l-v} diagram.

We note that we don't include feedback, magnetic fields or gas self-gravity in the simulations presented here, in order to speed up computation. We believe that the feedback in particular will disperse the emission in the \emph{l-v} maps as seen in \cite[Acreman et al. (2012)]{Acreman_etal2012}. The effect of these extra physical processes on \emph{l-v} features will be the subject of a future investigation.


\begin{discussion}

\discuss{Antoja}{Do you have a preferred picture so far of the spiral/bar structure?}

\discuss{Pettitt}{It's still early days and I would prefer not speculate too much yet, there's still a large parameter space still to search.}

\discuss{Binney}{If you cut a corner by assuming HI is optically thin, how wrong is the data cube you produce?}

\discuss{Pettitt}{That's a good question, and I haven't looked into this yet. However because the CO distribution is more useful in determining spiral/bar structures we are focussing our attention on developing these.}

\end{discussion}

\begin{thebibliography}{}

\bibitem[{Acreman} et~al.(2010){Acreman}, {Douglas}, {Dobbs}, and
  {Brunt}]{Acreman_etal10}
{Acreman}, D.~M., {Douglas}, K.~A., {Dobbs}, C.~L. and {Brunt}, C.~M., 2010, \textit{MNRAS}, 406, 1460

\bibitem[{Acreman} et al. \ (2012){Acreman}, {Dobbs}, {Brunt}, and
  {Douglas}]{Acreman_etal2012}
{Acreman}, D.~M., {Dobbs}, C.~L., {Brunt}, C.~M. and {Douglas} K.~A. 2012, \textit{MNRAS}, 422, 241

\bibitem[{Baba} et~al.(2010){Baba}, {Saitoh}, and {Wada}]{Baba_etal2010}
{Baba}, J., {Saitoh}, T.~R. and {Wada}, K., 2010, \textit{PASJ}, 62, 1413

\bibitem[{Bate} et~al.(1995){Bate}, {Bonnell}, and
  {Price}]{Bate_etal95}
{Bate}, M.~R., {Bonnell}, I.~A., and {Price}, N.~M.,1995, \textit{MNRAS}, 277, 362

\bibitem[{Benz} et~al.(1990){Benz}, {Cameron}, {Press}, and
  {Bowers}]{Benz_etal90}
{Benz}, W., {Cameron}, A.~G.~W., {Press}, W.~H., and {Bowers}, R.~L., 1990, \textit{ApJ}, 348, 647

\bibitem[{Cox} and {G{\'o}mez}(2002)]{Cox02}
{Cox}, D.~P. and {G{\'o}mez}, G.~C. 2002,
\textit{ApJS}, 142, 261

\bibitem[Dame et al. \ (2001)]{Dame_etal01}
{Dame, T.M., Hartmann, D., and Thaddeus, P.,} 2001,
\textit{ApJ}, 547, 792 

\bibitem[{Dehnen}(2000)]{Dehen00}
Dehnen, W., 2000, \textit{ApJ}, 119, 800

\bibitem[{Dobbs et al. \ }(2008)]{Dobbs_etal08}
Dobbs, C.~L.,  Glover, S. C. O., Clark,  P. C. and Klessen, R. S. 2008, \textit{MNRAS}, 389, 1097

\bibitem[{Glover} and {Mac Low}(2007)]{GloverMLow_07}
{Glover}, S.~C.~O. and {Mac Low}, M.-M., 2007, 169, 239

\bibitem[{G{\'o}mez} and {Cox}(2004)]{GomezCox08}
{G{\'o}mez}, G.~C. and {Cox}, D.~P. 2004, \textit{ApJ}, 615, 758

\bibitem[{Harries(2000)}]{Harries00}
Harries, T.~J. 2000, \textit{MNRAS}, 315, 722

\bibitem[{Hernquist}(1993)]{Hernquist93}
{Hernquist}, L., 1993, \textit{ApJs}, 86, 1993

\bibitem[Kalberla et al. \ (2005)]{Kalberla_etal05}
{Kalberla, P.~M.~W., Burton, W.~B., Hartmann, D., Arnal, E.M., Bajaja, E., Morras, R., \& Poppel, W. G. L.} 2001,
\textit{A\&A}, 440, 775 

\bibitem[{Lodato} and {Price}(2010)]{LodatoPrice10}
{Lodato}, G. and {Price}, D.~J., 2010, \textit{MNRAS}, 405, 1212

\bibitem[{Long} and {Murali}(1992)]{LongMurali92}
Long, K. and {Murali}, C. 1992, \textit{ApJ}, 397, 44

\bibitem[Martos et al. \ (2004)]{Martos_etal05}
Martos, M., Hernandez, X. Y{\'a}{\~n}ez, M., Moreno, E., and
  Pichardo, B., 2005, \textit{MNRAS}, 350, L47

\bibitem[{Nelson} and {Langer}(1997)]{NelsonLanger97}
{Nelson}, R.~P. and {Langer}, W.~D., 1997, \textit{ApJ}, 482, 796

\bibitem[{Price} and {Federrath}(2010)]{PriceFed10}
{Price}, D.~J. and {Federrath}, C., 2010, \textit{MNRAS}, 406, 2010

\bibitem[{Price} and {Monaghan}(2007)]{PriceMon07}
{Price}, D.~J. and {Monaghan}, J.~J., 2007, \textit{MNRAS}, 374, 1347

\bibitem[{Rodriguez-Fernandez, N. J.} and {Combes, F.}(2008)]{RFRC08}
{Rodriguez-Fernandez, N. J.} and {Combes, F.} 2008, \textit{A\&A}, 489,115

\bibitem[{Vall{\'e}e}(2008)]{Vallee08}
{Vall{\'e}e}, J.~P., 2008, \textit{AJ}, 135, 1301.

\bibitem[{Wada} and {Koda}(2001)]{WadaKoda01}
{Wada}, K., and {Koda}, J., 2001, \textit{PASJ}, 53, 1163

\bibitem[{Wang} et~al.(2012){Wang}, {Zhao}, {Mao}, and
  {Rich}]{Wang_etal12}
{Wang}, Y., {Zhao}, H., {Mao}, S., and {Rich}, R.~M. , 2012, \textit{MNRAS}, 427, 1429

\end{thebibliography}
\end{document}